\documentclass[11pt]{article}
\usepackage{Moriond,epsfig,subfigure}

\def\Journal#1#2#3#4{{#1} {\bf #2}, #3 (#4)}

\def\PLB{{\em Phys. Lett.}  B}

\def\PRD{{\em Phys. Rev.} D}

\def\EPJC{{\em Eur. Phys. J.} C}

\def\be{\begin{equation}}
  \def\ee{\end{equation}}
\def\bea{\begin{eqnarray}}
  \def\eea{\end{eqnarray}}

\begin{document}
\vspace*{4cm}
\title{Hadronic cross section measurements and contribution to $(g-2)_\mu$ with KLOE}

\author{P. Beltrame for the KLOE collaboration} 
\address{
  At present: Marie Curie fellow at CERN PH Department \\
  CH-1211 Genève 23, Switzerland \\
  e-mail: \texttt{paolo.beltrame@cern.ch}
}

\maketitle\abstracts{The KLOE experiment at the DA$\Phi$NE $\phi$-factory has performed a new precise measurement of the pion form factor using Initial State Radiation events. Results based on an integrated luminosity of 240 pb$^{-1}$ and extraction of the $\pi\pi$ contribution to $a_\mu$ in the mass range $0.35< M^2_{\pi\pi}<0.95$ GeV$^2$ are presented. The new value of  $a^{\pi\pi}_\mu$ has smaller statistical and systematic error and is consistent with the KLOE published value (confirming the current disagreement between the Standard Model prediction for $a_\mu$ and the measured value).}

\section{Introduction}
\label{sec:intro}

The anomalous magnetic moment of the muon has been measured with an accuracy of 0.54 ppm.~\cite{benn} The main source of uncertainty in the value predicted by the Standard Model is given by the hadronic contribution, $a_\mu^{hlo}$, to the lowest order. This quantity can be evaluated via a dispersion integral of the hadronic cross section measurements. The pion form factor $F_\pi$ (proportional to the $\sigma_{\pi\pi}$ cross section) accounts for $\sim70\%$ of the central value and for $\sim60\%$ of the uncertainty in $a_\mu^{hlo}$. The KLOE experiment has already published a measurement of $|F_\pi|^2$ using Initial State Radiation (ISR) events, based on 140 pb$^{-1}$ data taken in 2001,~\cite{kloe05,kloe05up} with a fractional systematic error of $1.3\%$.

\section{Measurement of $\sigma(e^+e^-\to\pi^+\pi^-\gamma)$ at DA$\Phi$NE}
\label{sec:meas}

DA$\Phi$NE is an $e^+ e^-$ collider running at $\sqrt{s}\simeq m_\phi$, the $\phi$ meson mass, which has provided an integrated luminosity of about 2.5 fb$^{-1}$ to the KLOE experiment. In addition, during the year 2006, about 230 pb$^{-1}$ of data have been collected at $\sqrt{s}\simeq 1$ GeV. The results shown in this contribution are based on 240 pb$^{-1}$ of data taken in 2002 (3.1 Million events).~\cite{kloe08} A preliminary spectrum based on the {\it off peak} data sample will be also given.~\cite{pop}\\
The KLOE detector consists of a drift chamber with excellent momentum resolution ($\sigma_p/p\sim 0.4\%$ for tracks with polar angle larger than $45^\circ$) and an electromagnetic calorimeter with good energy ($\sigma_E/E\sim 5.7\%/\sqrt{E~(\rm{GeV})}$) and precise time ($\sigma_t\sim 54~\rm{ps}/\sqrt{E~(\rm{GeV})}\oplus 100~\rm{ps}$) resolution.\\
At DA$\Phi$NE, we measured the differential spectrum of the $\pi^+\pi^-$ invariant mass, $M_{\pi\pi}$, from ISR events, $e^+ e^-\to\pi^+\pi^-\gamma$, and the total cross section $\sigma_{\pi\pi}\equiv\sigma_{e^+ e^-\to\pi^+\pi^-}$ is obtained using the following formula:~\cite{binn99}
\begin{equation}
  s~ \frac{{\rm d} \sigma_{\pi\pi\gamma}}
  {{\rm d} M_{\pi\pi}^2} = \sigma_{\pi\pi}
  (M_{\pi\pi}^2)~ H(M_{\pi\pi}^2)~,
  \label{eq:sppgtospp}
\end{equation}
where $H$ is the radiator function, describing the photon emission at the initial state. This formula neglects Final State Radiation (FSR) terms (which are however properly taken into account in the analysis).

\begin{figure}[h!]
  \begin{center}
    \subfigure
    {\includegraphics[width=10.pc]{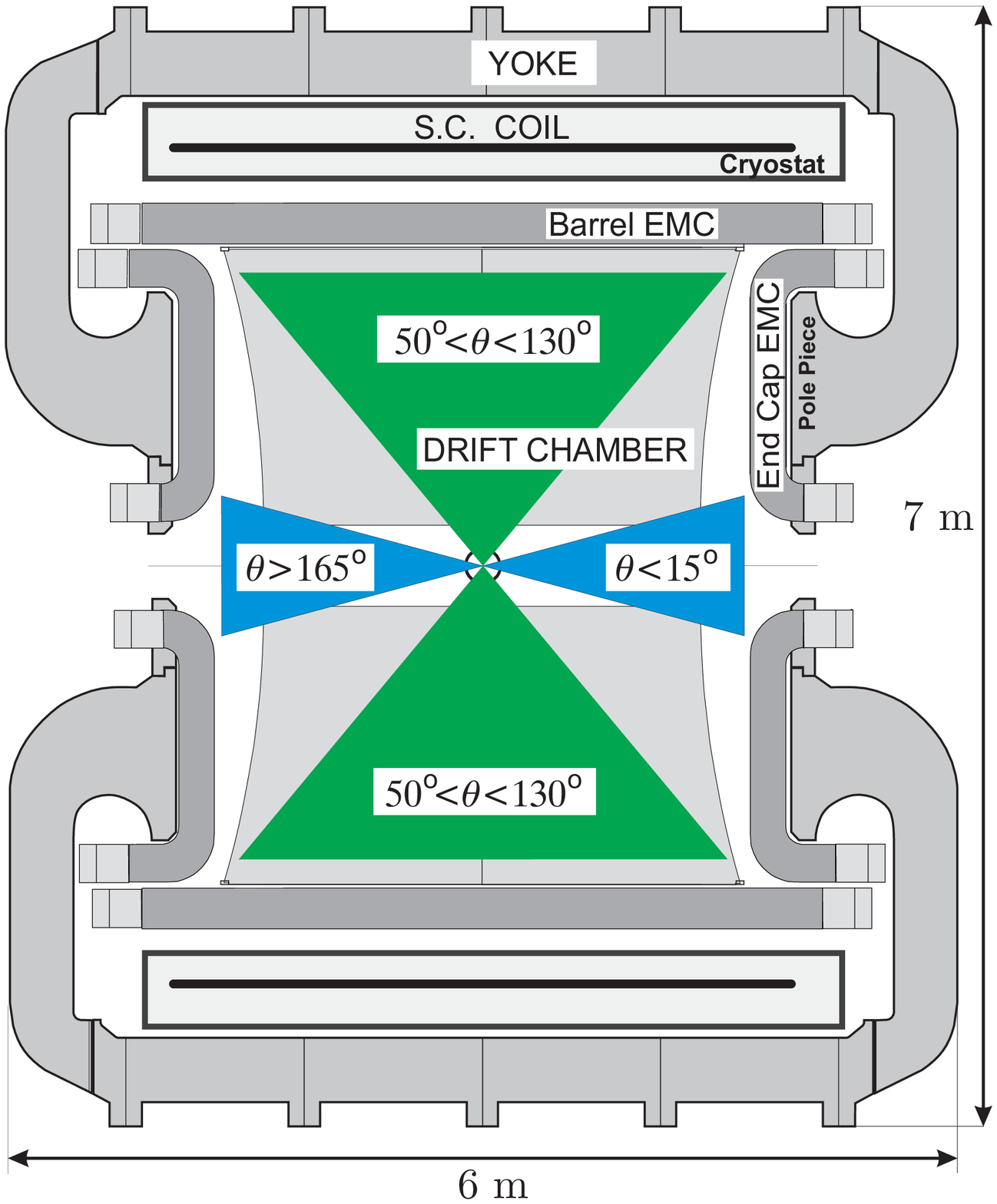}}
    \hglue 10 mm
    \subfigure
    {\includegraphics[width=12.pc]{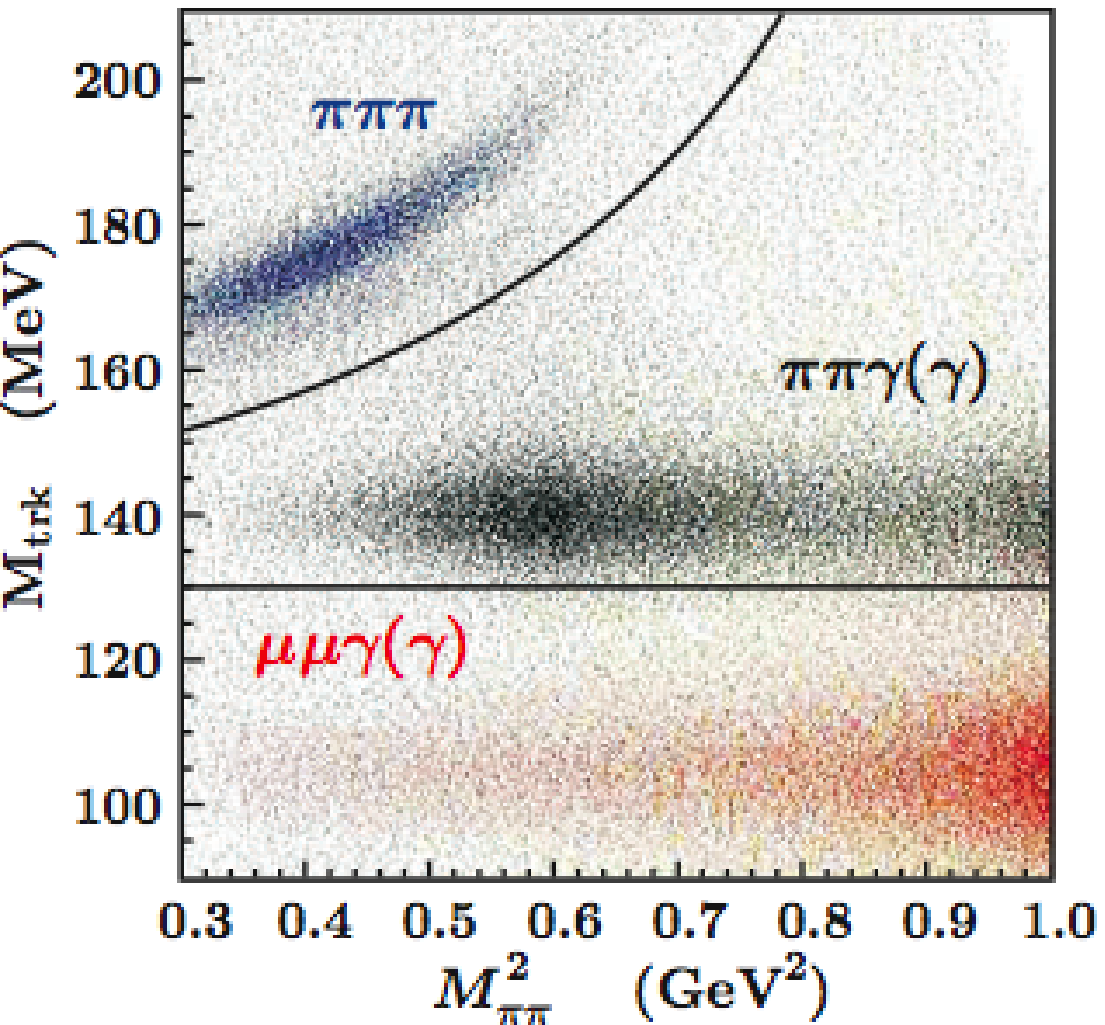}}
    \caption{\label{fig:kloe} Left: Fiducial volume for the small angle photon (narrow cones) and for the the pion tracks (wide cones). Right: Signal and background distributions in the $M_{\rm Trk}$-$M^2_{\pi\pi}$ plane; the selected area is shown.}
  \end{center}
\end{figure}
\noindent
In the {\it small angle} analysis, photons are emitted within a cone of $\theta_\gamma<15^\circ$ around the 
beam line (narrow blue cones in Fig.~\ref{fig:kloe} left). The two charged pion tracks have $50^\circ<\theta_\pi<130^\circ$. The photon is not explicitly detected and its direction is reconstructed by closing the kinematics: $\vec{p}_\gamma\simeq\vec{p}_{miss}= -(\vec{p}_{\pi^+}+\vec{p}_{\pi^-})$. The separation of pion and photon selection regions greatly reduces the contamination from the resonant process $e^+e^-\to \phi\to\pi^+\pi^-\pi^0$, in which the $\pi^0$ mimics the missing momentum of the photon(s) and from the final state 
radiation process $e^+e^-\to \pi^+\pi^-\gamma_{\rm FSR}$. Since ISR-photons are mostly collinear with the beam line, a high statistics for the ISR signal events remains. On the other hand, a highly energetic photon emitted at small angle forces the pions also to be at small angles (and thus outside the selection cuts), resulting in a kinematical suppression of events with $M^2_{\pi\pi}< 0.35$ GeV$^2$. Residual contamination from the processes $\phi\to\pi^+\pi^-\pi^0$
and $e^+e^- \to\mu^+\mu^-\gamma$ are rejected by cuts in the kinematical variable {\it trackmass},~\footnote{Defined under the hypothesis that the final state consists of two charged particles with equal mass $M_{\rm Trk}$ and one photon.} see Fig.~\ref{fig:kloe} right. A particle ID estimator, based on calorimeter information and time-of-flight, is used to suppress the high rate of radiative Bhabhas.     

\section{Evaluation of $|F_\pi|^2$ and $a_\mu^{\pi\pi}$}
\label{sec:fpi}

The $\pi\pi\gamma$ differential cross section is obtained from the observed spectrum, $N_{obs}$, after subtracting the residual background events, $N_{bkg}$, and correcting for the selection efficiency, $\varepsilon_{sel}(M_{\pi\pi}^2)$,
and the luminosity:
\begin{equation}
  \frac{{\rm d} \sigma_{\pi\pi\gamma}}
  {{\rm d} M_{\pi\pi}^2} = \frac{N_{obs}-N_{bkg}}
  {\Delta M_{\pi\pi}^2}\, \frac{1}{\varepsilon_{sel}(M_{\pi\pi}^2)~ \mathcal{L}}~ ,
  \label{eq:sppg}
\end{equation}
where the observed events are selected in bins of $\Delta M_{\pi\pi}^2=0.01$ GeV$^2$. The residual background content is found by fitting the $M_{\rm Trk}$ spectrum of the selected data sample with a superposition of Monte Carlo distributions describing the signal and background sources. The radiator function $H$ used to get $\sigma_{\pi\pi}$ in Eq.~\ref{eq:sppg} is taken from the \texttt{PHOKHARA} Monte Carlo generator, which calculates the complete next-to-leading order ISR effects.~\cite{czyz03} In addition, the cross section is corrected for the vacuum polarisation (running of $\alpha_{\mathrm em}$),~\cite{jeg06} and the shift between the measured value of $M^2_{\pi\pi}$ and the squared virtual photon mass $M^2_{\gamma^*}$ for events with photons from final state radiation. Again the \texttt{PHOKHARA} generator, which includes FSR effects in the pointlike-pions approximation, is used to estimate the latter.~\cite{czyz05} \\
The cross section corrected for the above effects and inclusive of FSR, $\sigma_{\pi\pi}^{bare}$ 
(shown in Fig.~\ref{fig:spp}), is used to determine $a_\mu^{\pi\pi}$ via a dispersion integral:
\begin{equation}
  a_\mu^{\pi\pi} = \frac{1}{4\pi^3}\int_{s_{min}}^{s_{max}}{\rm d} s~\sigma_{\pi\pi}^{bare}(s)\,K(s)~ ,
  \label{eq:dispin}
\end{equation}
where the lower and upper bounds of the spectrum measured in this analysis are $s_{min}=0.35$ GeV$^2$ and $s_{max}=0.95$ GeV$^2$, and $K(s)$ is the kernel function.~\cite{brod67}
Tab.~\ref{tab:amu} left shows the list of fractional systematic uncertainties of $a_\mu^{\pi\pi}$ in the mass range $0.35< M^2_{\pi\pi}<0.95$ GeV$^2$.
\begin{table}[h!]
  \begin{center}
    \begin{tabular}{cc}
      \parbox[c]{180pt}{
        {\scriptsize
          \begin{tabular}{|l|c|}
            \hline
            Reconstruction Filter & negligible\\
            Background subtraction & 0.3 \% \\
            Trackmass/Miss. Mass & 0.2 \% \\
            $\pi$/e-ID & negligible\\
            Tracking & 0.3 \% \\
            Trigger & 0.1 \% \\
            Unfolding & negligible \\
            Acceptance ($\theta_{\rm miss}$) & 0.2 \% \\
            Acceptance ($\theta_\pi$) & negligible \\
            Software Trigger (L3) & 0.1 \% \\
            Luminosity ($0.1_{th}\oplus 0.3_{exp}$)\% & 0.3 \% \\
            $\sqrt{s}$ dependence of $H$ & 0.2 \%\\
            \hline
            Total experimental systematics & 0.6 \% \\
            \hline
            \hline
            Vacuum Polarization &  0.1 \% \\
            FSR resummation & 0.3 \% \\
            Rad. function $H$   & 0.5 \% \\
            \hline
            Total theory systematics & 0.6 \% \\
            \hline
          \end{tabular}
        }
      }
      &
      \parbox[c]{180pt}{
        \renewcommand{\arraystretch}{1.6}
        \setlength{\tabcolsep}{1.2mm}
        {\scriptsize
          \begin{tabular}{|l|c|}
            \hline
            \multicolumn{2}{|c|}{$a_\mu^{\pi\pi}\times10^{10} ~ 0.35<M_{\pi\pi}^2<0.95$ GeV$^2$ -- KLOE} \\
            \hline
            KLOE05~\cite{kloe05up}  & $384.4~\pm~0.8_{\rm stat}~\pm~4.6_{\rm sys}$ \\
            \hline
            KLOE08~\cite{kloe08} & $387.2~\pm~0.5_{\rm stat}~\pm~3.3_{\rm sys}$ \\
            \hline
            \hline
            \multicolumn{2}{|c|}{$a_\mu^{\pi\pi}\times10^{10} ~ 0.630<M_{\pi\pi}<0.958$ GeV} \\
            \hline
            CMD-2~\cite{cmd2} & $361.5\pm 5.1$ \\
            \hline
            SND~\cite{snd} & $361.0\pm 3.4$ \\
            \hline
            KLOE08~\cite{kloe08} & $356.7\pm 3.1$ \\
            \hline
          \end{tabular}
        }
      }
    \end{tabular}
    \caption{Left: Systematic errors on the extraction of $a_\mu^{\pi\pi}$ in the mass range $0.35<M^2_{\pi\pi}<0.95$ GeV$^2$. Right: Comparison among $a_\mu^{\pi\pi}$ values.}
    \label{tab:amu}
  \end{center}
\end{table}
 
\section{Results}
\label{sec:results}

\begin{figure}[h!]
  \begin{center}
    \subfigure
    {\includegraphics[width=12.pc]{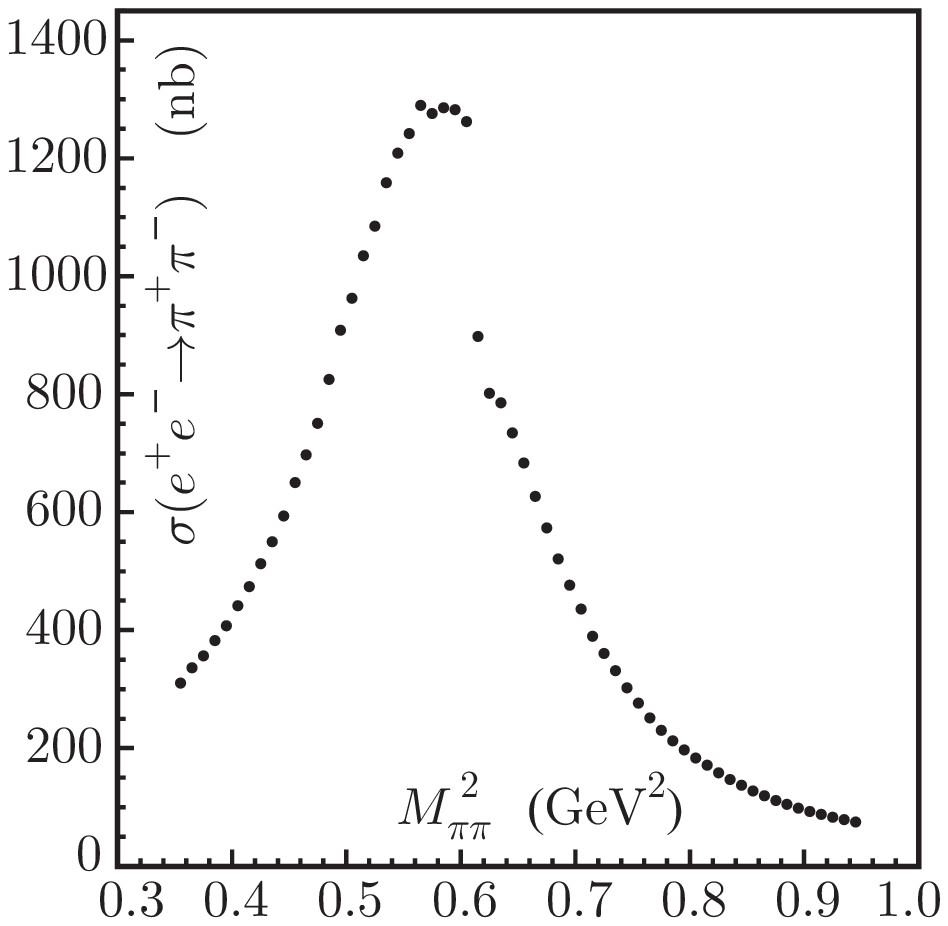}}
    \label{fig:spp}
    \hglue 10 mm
    \subfigure
    {\includegraphics[width=10.pc,angle=90]{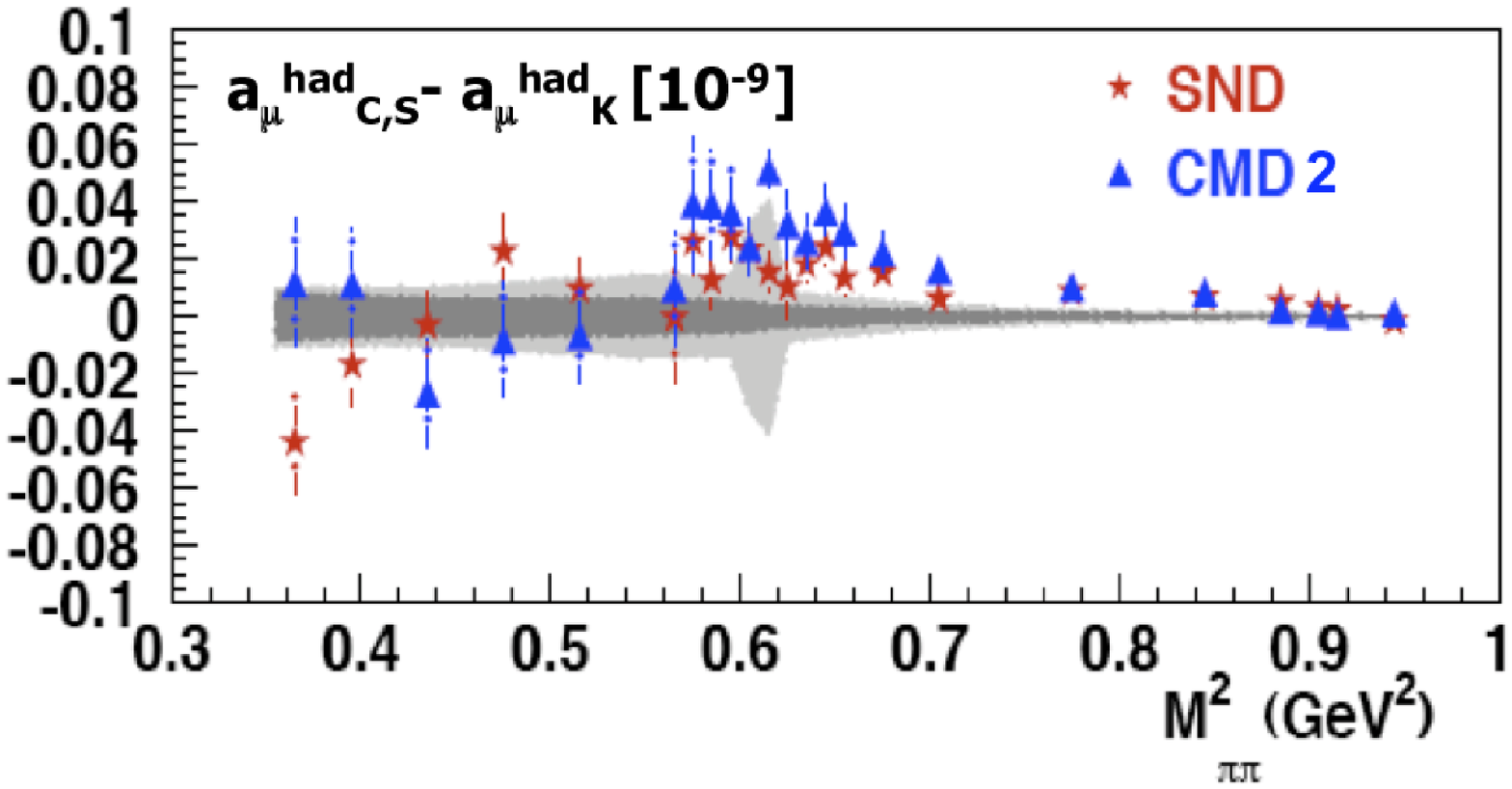}}
    \label{fig:damu}
    \caption{Left: Differential cross section for $e^+e^-\to\pi^+\pi^-$, with $|\cos\theta_\gamma|>\cos(15^\circ)$. Right: Absolute difference between the dispersion integral value in each energy bin evaluated by CMD-2, SND and KLOE, where for this latter the statistical errors (light band) and summed statistical and systematic errors (dark band) are shown.}
  \end{center}
\end{figure}

The new KLOE analysis (KLOE08) presented here is compared with the previous published one (KLOE05) based on the 2001 data sample,~\cite{kloe05up} and with the results from the VEPP-2M experiments,~\cite{snd,cmd2} in the mass range $0.630<M_{\pi\pi}<0.958 {\rm GeV}$. Tab.~\ref{tab:amu} right shows the consistency between the KLOE results, and with the CMD-2 and SND values. Fig.~\ref{fig:damu} shows the absolute difference between the $a_\mu^{\pi\pi}$ values for each energy bin obtained in this analysis and the energy scan experiments. All the experiments are in agreement within errors.

\section{Conclusions and outlook}

\begin{figure}[h!]
  \begin{center}
    \includegraphics[width=13.pc]{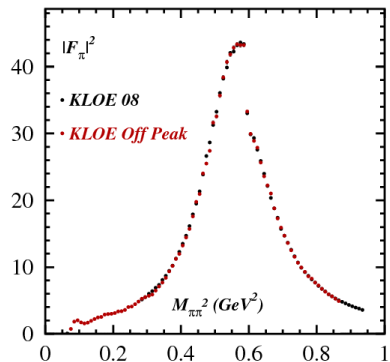}
    \label{fig:fpi06}
    \caption{Preliminary $|F_\pi|^2$ result obtained with off peak data superimposed to the published result.}
  \end{center}
\end{figure}

We have measured the di-pion contribution to the muon anomaly, $a_\mu^{\pi\pi}$, in the interval $0.592 < M_{\pi\pi} < 0.975$ GeV, with negligible statistical error and with an experimental systematic uncertainty of 0.6\%. Taking also into account the other 0.6\% uncertainty due to the theoretical calculations of the radiative corrections, we find:
$$a_\mu^{\pi\pi}(0.592 < M_{\pi\pi} < 0.975 ~ {\rm GeV})=(387.2\pm 3.3)\times 10^{-10}~.$$
This result represents an improvement of 30\% on the systematic error with respect to our previous published value. The new result confirms the current disagreement between the Standard Model prediction for $a_\mu$ and the direct measured value of $a_\mu$.\\
Independent analyses are in progress to: ({\it i}) measure $\sigma_{\pi\pi}$ using detected photons emitted at large angle, which would improve the knowledge of the FSR interference effects (in particular the $f_0(980)$ contribution); ({\it ii}) measure the pion form factor directly from the ratio, bin-by-bin, of $\pi^+\pi^-\gamma$ to $\mu^+\mu^-\gamma$ spectra;~\cite{ppg_mmg} ({\it iii}) extract the pion form factor from data taken at $\sqrt{s}=1$ GeV, off the $\phi$ resonance, where $\pi^+\pi^-\pi^0$ background is negligible.~\cite{pop} The preliminary $|F_\pi|^2$ result, superimposed on the published result, is shown in Fig.~\ref{fig:fpi06}, where it is possible to see the agreement between the two spectra.

\section*{References}

\end{document}